\documentclass[preprint2]{aastex}

\shorttitle{Evidence for outflow-ambient interfaces?}
\shortauthors{Codella et al.}

\begin{document}


\title{First evidence for molecular interfaces between outflows and
ambient cloud in high-mass star forming regions?}


\author{C. Codella}
\affil{INAF - Istituto di Radioastronomia, Sezione di Firenze, Largo E. Fermi 5, 50137
Firenze, Italy}
\email{codella@arcetri.astro.it}

\author{S. Viti and D.A. Williams}
\affil{Department of Physics and Astronomy, University College London, Gower Street, London,
WC1E6BT}
\email{sv@star.ucl.ac.uk, daw@star.ucl.ac.uk}

\and

\author{R. Bachiller\altaffilmark{3}}
\affil{Observatorio Astron\'omico Nacional (IGN), Alfonso XII 3, E-28014 Madrid, Spain}
\email{bachiller@oan.es}




\begin{abstract}
We present new observations of the CepA-East region of massive star formation
and describe an extended and dynamically distinct feature not previously
recognised. This feature is present in emission from H$_2$CS, OCS, CH$_3$OH, and
HDO at --5.5 km s$^{-1}$, but is not traced by conventional tracers of star
forming regions H$_2$S, SO$_2$, SO, CS. The feature is extended up to at least
0.1 pc. We show that the feature is neither a hot core nor a shocked
outflow. However, the chemistry of the feature is consistent with
predictions of a model of an eroding interface between a fast wind and a
dense core; mixing between the two media occurs in the interface 
on a timescale of 10-50 years. If these
observations are confirmed by detailed maps and by detections in species
also predicted to be abundant (e.g. HCO$^+$, H$_2$CO, and NH$_3$) this feature
would be the first detection of such an interface in regions of massive
star formation. An important implication of the model is that a
significant reservoir of sulfur in grain mantles is required to be in the
form of OCS.

\end{abstract}





\section{Introduction}

To explore the interactions of young stellar objects (YSOs) with their
environments, we recently carried out mm-wavelength molecular line
observations towards star forming regions (SFRs) with well defined and
bright high-velocity components. In particular, we mapped the well known
Cepheus A \citep[CepA;][and references therein]{garay96} SFR in several
shock-chemistry tracers such as H$_2$S, SO$_2$, and HDO
\citep{code03,code05}. Those results show that the group of B-type
stars located in CepA-East producing a hot core \citep{jesus05}, 
are also associated with multiple mass loss processes.
In particular, beside the already known three flows pointing in the
SW, NE, and SE directions, a fourth outflow flowing towards the
South has been detected thanks to the shock-chemistry tracers.
CepA-East can thus be considered an ideal laboratory
in which to study how outflow motions affect the gas, from both the
kinematical and chemical points of view. \citet{code05} have
already presented a multi-species and multi-line mm-survey of the
central region of CepA-East where the YSOs are located. Using the 30-m
IRAM antenna, the authors detected emission in different
transitions of 21 molecular species tracing a wide range of physical
conditions. Analysis of these spectra shows that different molecules
exhibit different spectral behaviours and that three classes can be
distinguished: (i) hot core molecules (e.g. HC$^{18}$O$^+$,
CH$_3$C$_2$H) emitting only at the velocity of the hot core (--10.7 km
s$^{-1}$) and with no line wings, (ii) outflow molecules (e.g. CS,
SiO, H$_2$S, SO$_2$, and SO) spanning the whole range of observed 
outflowing velocities so that bright wings are added to the hot core
emission, and (iii) four species (OCS, H$_2$CS, HDO, and CH$_3$OH)
which are associated with wings and which, in addition, clearly show a
redshifted spectral peak at --5.5 km s$^{-1}$, well separated from the
hot core peak. While the peak at --10.7 km s$^{-1}$ is tracing the
high-density material hosting the YSOs and the wings are tracing the
multiple outflows, the origin of the redshifted spectral peak is
unknown. The \citet{code05} data did not allow us to clarify the
spatial distribution of this spectral peak and to establish if it is
tracing a small structure or it is related with an extended
component. 
It is worth noting that, as far as we know, this is the first
study to reveal duality in the line-wing profiles observed in outflows
driven by YSOs, i.e that OCS, H$_2$CS, HDO, and CH$_3$OH (hereafter
called double-peaked species) have a different behaviour with respect to
CS, SiO, H$_2$S, SO$_2$, and SO (hereafter called single-peaked
species). This suggests that the redshifted spectral peak could
be tracing a different gas component with respect to the gas
outflowing at the other velocities as well as to the hot core gas, and
indicates that high angular resolution observations are required for a
detailed analysis.

In this Letter we present observations which clarify the spatial
distribution of the redshifted spectral peak at --5.5 km s$^{-1}$. In
addition, we produce chemical models suggesting that we
are probably observing the first direct evidence of turbulent interfaces 
(i) where the outflow gas impinges on and detach dense gas, and 
(ii) in which
dynamical mixing and diffusion are occurring between the mass ejected
from a newly-formed massive YSO and the interstellar cloud from which
it was formed. In the following, the kinematical component at --5.5 km s$^{-1}$
will be referred to as the I-feature.

\section{The spatial distribution of the I-feature}

The main goal of the observations was to determine the spatial distribution of
the I-feature observed towards CepA-East. In order to select
the best molecular tracers, following \citet{code05}, we noted that a good compromise between high 
angular resolution, a simple spectral pattern, 
and an intense line emission was represented by H$_2$CS, and in
particular by its 6$_{\rm 16}$--5$_{\rm 15}$ transition at 202923.55 MHz
($E_{\rm u}$=47 K). Thus, we mapped
a region of $\sim$ 1$\arcmin$$\times$2$\arcmin$ in 
H$_2$CS(6$_{\rm 16}$--5$_{\rm 15}$) on 2004 June 
with the 30-m IRAM radiotelescope of Pico Veleta (Spain). We used a sampling of
10$\arcsec$ around the coordinates of HW2 YSO, which is thought to be among the main drivers of the
CepA-East outflows and it represents the center of
the present map, whereas a spacing of 20$\arcsec$ was chosen to scan coarsely the outer
part of the cloud. The system temperature, 
$T_{\rm sys}$, was $\sim$ 500 K, the HPBW was 12$\arcsec$, while the pointing
was accurate to within 2$\arcsec$-3$\arcsec$. As spectrometer, we
used an autocorrelator with a configuration providing a velocity
resolution of 0.06 km s$^{-1}$, successively smoothed to 0.92 km s$^{-1}$.
The spectra were calibrated with the standard chopper wheel method and reported 
here in units of main-beam brightness temperature ($T_{\rm MB}$): the average r.m.s.
is $\sim$ 20 mK.

Figure 1{\it a} reports the velocity channel maps of the
H$_2$CS(6$_{\rm 16}$--5$_{\rm 15}$) emission. Each panel shows the
emission integrated over a velocity interval of 2 km s$^{-1}$; the
ambient velocity ($V_{\rm LSR}$) is --10.7 km s$^{-1}$
\citep[e.g.][]{code05}.
In summary, H$_2$CS(6$_{\rm 16}$--5$_{\rm 15}$) emission is associated (i)
with the central position, where the YSO cluster is located and where
a hot core has been detected \citep{jesus05}, and (ii) with the four
outflow directions, NE, SE, S, and SW, confirming H$_2$CS as a tracer
of high-temperature and/or shocked regions.
In particular, the new H$_2$CS maps confirm
that the spatial distribution of
the I-feature 
is not limited to the central position 
tracing also the SW and southern outflows, as clearly shown by the --5.5 and --3.5
km s$^{-1}$ panels in Fig. 1a.

Examples of H$_2$CS(6$_{\rm 16}$--5$_{\rm 15}$) line profiles are shown in
Fig. 1{\it b} which compares the spectra observed at the (0$\arcsec$,+10$\arcsec$) 
and (0$\arcsec$,--10$\arcsec$) offsets with that  
observed at the (0$\arcsec$,0$\arcsec$) position \citep{code05}.
Given the HPBW, the three spectra are sampling different regions of
CepA-East. The I-feature is still present 
in the southern position and it is even more redshifted, approaching
the --5 km s$^{-1}$ velocity, thus suggesting the presence of
a velocity gradient.
In order to make the same comparison with another tracer of the I-feature, 
we observed the 
CH$_3$OH(5$_{\rm K}$--4$_{\rm K}$) transition at 241.8 GHz towards the
(0$\arcsec$,+10$\arcsec$) and (0$\arcsec$,--10$\arcsec$) positions.
In this case, 
$T_{\rm sys}$ $\simeq$ 650 K, the HPBW = 10$\arcsec$, 
and the resulting spectra were smoothed to 0.39 km s$^{-1}$, with a r.m.s. $\sim$ 80 mK.
Fig. 1{\it c} compares the profiles due to the high
excitation ($E_{\rm u}$=115 K) CH$_3$OH(5$_{\rm 4}$--4$_{\rm 4}$ A$^{\pm}$)
transition at 241806.51 MHz, which according to the (0$\arcsec$,0$\arcsec$) spectrum, 
can be considered as the best tracer of the I-feature among the lines of the 
CH$_3$OH(5$_{\rm K}$--4$_{\rm K}$) pattern. 
Also in this case, the I-feature is still present in the
southern position and again it seems more redshifted, in agreement
with the H$_2$CS data. It is worth noting that the two continuous
lines across the methanol (0$\arcsec$,0$\arcsec$) spectrum refer to
the two velocity components of the CH$_3$OH(5$_{\rm -4}$--4$_{\rm -4}$
E) transition ($E_{\rm u}$=122 K): the --5.5 km s$^{-1}$ (5$_{\rm
-4}$--4$_{\rm -4}$ E) emission is blended with the --10.7 km s$^{-1}$
(5$_{\rm 4}$--4$_{\rm 4}$ A$^{\pm}$) feature, producing a broad line.
Finally, Fig. 1{\it d} reports the HDO(1$_{\rm 10}$--1$_{\rm 11}$)
($E_{\rm u}$=47 K) profiles that were also reported in the \citet{code05}
paper. Since the angular resolution is definitely higher
(HPBW=31$\arcsec$) we show the spectra taken in the
(0$\arcsec$,+20$\arcsec$) and (0$\arcsec$,--20$\arcsec$) positions to
sample different parts of CepA-East. The HDO data are
well in agreement with the picture given above, showing how bright is
the I-feature as observed towards the southern position.

We conclude that there is a previously undetected structure extending 
toward the South from the central position giving rise to emission at 
--5.5 km s$^{-1}$ in lines of H$_2$CS, OCS, HDO, and CH$_3$OH. 
The structure appears to be at least 0.1 pc in length. 





\section{The origin of the I-feature}

In this Section we consider possible origins for
the extended I-feature at --5.5 km s$^{-1}$.
Since CepA-East is a YSO cluster the possibility that this
emission comes from a second hot core seems, at first sight,
likely. However, this cannot be the case because: (i) 
the emission is extended and
{\it not} compact as it would be if it came from a hot core (see Fig. 1).
In addition, (ii) Figure 2 shows that what makes the I-feature peculiar is that some line ratios
such as OCS/SO, H$_2$CS/SO$_2$ or CH$_3$OH/H$_2$S are 
definitely much higher than those measured towards
the --10.7 km s$^{-1}$ component. In other words, although we cannot exlude that
conventional hot core tracers, including, SO, SO$_2$, and
H$_2$S, can have some weak emission at --5.5 km s$^{-1}$, 
they are {\it not} tracing the I-feature.
Finally, (iii) we find that the column densities of the double peaked species 
in the
I-feature (CH$_3$OH: 3 10$^{15}$ cm$^{-2}$; OCS: 5 10$^{13}$; H$_2$CS: 1 10$^{14}$;
HDO: 6 10$^{14}$; Codella et al. 2005), calculated
assuming extended emission, are lower that those typically
found in hot cores \citep{jacq,millar}.
We conclude that there are both morphological and chemical reasons
for excluding a hot core as the origin of the I-feature.

We have also investigated the possibility that the 
I-feature is simply the emission from a shocked molecular outflow. 
This possibility has been proposed in the \citet{code05} paper, where
the observations were compared with the theoretical calculations recently
reported by \citet{wake}. However, a closer inspection of such
model suggests that the theoretical abundances of the single-peaked
species are not reproduced and that both single and double peaked species
abundances, once converted into column densities (i.e. taking into
consideration the geometry) are more typical of hot cores densities and
hence too high.
In addition, it is difficult to understand why standard shock tracers like SiO. SO$_2$, 
and H$_2$S do not show the same profile observed with the double-peaked species,
as suggested by shock-chemistry models and
as happens in the well-studied chemical rich outflows BHR71 and L1157 \citep{garay98,bach}.

A new explanation is therefore required. We now investigate the
possibility that the I-feature arises from a molecular
interface between the outflow gas and the ambient medium.  
Highly turbulent interfaces should be present in all
environments where jets or outflows interact with the surrounding
molecular clouds. Such interfaces have been shown to be characterized
by a chemistry which is quite distinct in nature from that of typical
dense cores and would not be achieved by any modification of
conventional cold cloud chemistry. All chemical interface models,
whether with turbulent mixing or diffusion, explore the consequences
of mixing warm largely ionized gas with cold dense and mainly neutral
gas \citep[e.g.][]{lim,raw}. 
There is as yet no direct evidence of the existence of such interfaces
in high-mass SFRs, while in low-mass SFRs, 
observations of enhanced HCO$^+$ towards several Class 0 
objects can be interpreted as coming from the walls of the outflow
cavity \citep{hoger,raw00,redman}.
In order to test this hypothesis we
make use of the \citet{viti02} interface model and attempt to
reproduce the different trends displayed by the single-peaked and
double-peaked species. 
Note that since the model does not include the
deuterium chemistry, we could not calculate the abundance of HDO.

The main physical characteristics often adopted for an interface are:
low visual extinctions ($\le$ 1.5 mags); 
high gas densities ($n_{\rm H_{2}}$ $\simeq$ 10$^4$-10$^7$ cm$^{-3}$), 
high radiation fields compared to the mean interstellar radiation
field (here called one Habing),
short lifetimes ($\le$ 100 yr) and relatively high temperatures ($T_{\rm kin}$ $\ge$ 100 K)
due to non-dissociative shocks associated with the highly turbulent
interface between the outflow and the molecular core. Note that although
timescales may be very short, the continual erosion of dense material
by the wind or jet re-supplies the interface, so a near steady-state
pertains. 
The temperature and density of the region, as deduced from the
molecular observations \citep{code05}, are $\sim$ 200 K and
10$^6$--10$^7$ cm$^{-3}$, respectively.
We have therefore run a small grid of models using 
these values and varying 
the visual extinction (0.5-1.5 mags) and the radiation
fields (5-100 Habing). A further parameter that 
we investigate is the form that sulfur takes once depleted on the
grains, before the mantles are evaporated. We have considered 100\% in
solid H$_2$S, 100\% in solid OCS and several mixtures in between. Note
that while it is often assumed that most simple species, once frozen
out, react with hydrogen atoms to form saturated species, recently
doubts have been raised regarding the form that sulfur takes once it
depletes onto the grains. In particular, the detection of solid OCS
\citep{palumbo} and the non-detection of H$_2$S ices \citep{vandish} 
have raised the possibility that OCS may be
the main reservoir of sulfur on the grains. It is not unrealistic to
assume that before evaporation most of the sulfur is in the form of
OCS: studies of the effects of UV on grains 
\citep[e.g.][]{mannella} have shown that
species such as H$_2$O for example, are easily dissociated by
radiation; once the oxygen is freed, it is believed to react with CO,
which is the third most abundant species on the grains, to form
CO$_2$. It is not unreasonable to assume that a similar process occurs with 
H$_2$S: i.e. that once atomic sulfur is freed it reacts efficiently with CO
to form OCS.

The aim of this exercise is not to match perfectly
the observed column densities but to see whether there is an epoch
when a molecular outflow-ambient interface could give rise to high abundances of
CH$_3$OH, OCS and H$_2$CS while SO, SO$_2$ and H$_2$S remained low.
Figure 3 shows one of our best fit models, where, in general, the behaviour is as
expected: the double-peaked species show a clear anti-correlation with
respect to the single-peaked family. Note that in any case the molecular
abundances of all the species 
ultimately ($\ge$ 100 yr) decline as the radiation field induces
photodissociation, but, as mentioned earlier, the
interface is turbulent and therefore a continuous erosion of material will
constantly replenish the interface gas.  
A radiation field larger than 10 Habing seems to
destroy most of the doubly-peaked species too early. A density lower
than 10$^6$ cm$^{-3}$ does not produce enough methanol. But the most
interesting constraint is the amount of sulfur (at least 90\%) that
needs to be in solid OCS before evaporation occurs.
Thus, these
calculations suggest that the observations are consistent with
(but do not prove) a model of an outflow-ambient turbulent interface  
with the following physical conditions: $n_{\rm H_{2}}$
$\simeq$ 10$^6$--10$^7$ cm$^{-3}$, $T_{\rm kin}$ $\sim$ 200 K, and a
radiation field between 5 and 10 Habing.
According to this modelling, our observational data may be evidence
of the presence of a molecular interface in high mass star forming
regions. As well as indicating high abundance of the double-peaked species, our
model:

\begin{enumerate}

\item requires high abundances of solid OCS, and 
an entrainment of sufficient material into the interface on timescales of
10-50 years.
While it is feasible that OCS on ices is
abundant in an environment where the UV radiation field is strong, it
is clear that experimental as well as observational studies on ices
are desirable to confirm such picture;

\item
predicts high abundances of other species, including
HCO$^+$, H$_2$CO, and NH$_3$. To confirm the presence of a
molecular interface between outflow and ambient gas, 
we thus suggest further observations to map CepA-East in these species as well as
OCS lines. 

\end{enumerate}

\acknowledgments

SV acknowledges financial support from an individual PPARC advanced fellowship.

\clearpage



\begin{figure*}
\includegraphics[angle=-90,scale=0.65]{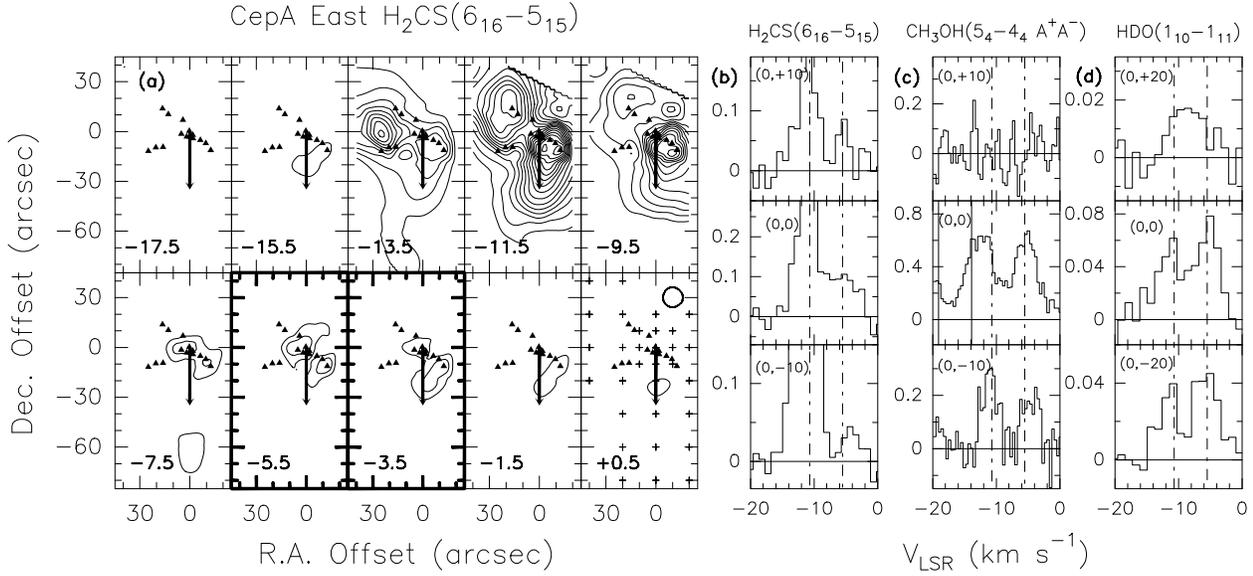}
\caption{({\it a}) Channel map of the H$_2$CS(6$_{\rm 16}$--5$_{\rm 15}$) emission
towards CepA East. Each panel shows the emission integrated over a velocity
interval of 2 km s$^{-1}$ centred at the value given in the bottom left corner. 
The thick boxes point out the main velocities of the I-feature (see text).
The position offsets are relative to $\alpha_{\rm 2000}$ = 22$^{\rm h}$ 56$^{\rm m}$ 17$\fs$9,
$\delta_{\rm 2000}$ = +62$\degr$ 01$\arcmin$ 49$\farcs$7, which are the HW2 coordinates.
The empty circle shows the IRAM 30-m beam (HPBW; 12$\arcsec$), 
while the small crosses mark the observed positions.
The triangles stand for the VLA 2 cm continuum components which trace two strings of
sources arising in shocks \citep{garay96}. The black arrow points out
the direction of the 0.6 pc collimated H$_2$S outflow discovered by \citet{code03}.
The contours range from 80 to 1120 
mK km s$^{-1}$. The first contour and the step correspond to about 4 $\sigma$
(where $\sigma$ is the r.m.s. of the map). ({\it b}) and ({\it c}): 
H$_2$CS(6$_{\rm 16}$--5$_{\rm 15}$) ($E_{\rm u}$=47 K) and
CH$_3$OH(5$_{\rm 4}$--4$_{\rm 4}$ A$^{\pm}$) ($E_{\rm u}$=115 K) spectra 
in $T_{\rm MB}$ scale observed 
at the (0$\arcsec$,+10$\arcsec$) and (0$\arcsec$,--10$\arcsec$) offsets and compared
with those observed at the (0$\arcsec$,0$\arcsec$) position \citep{code05}.
The HPBWs of the CH$_3$OH and H$_2$CS observations 
are 10$\arcsec$ and 12$\arcsec$, respectively. The H$_2$CS spectra
have been truncated to stress the I-feature, 
pointed out by a dashed line as well as the ambient
velocity, --10.7 km s$^{-1}$.
The two continuous line in the (0$\arcsec$,0$\arcsec$) CH$_3$OH panel
refer to the two velocity components of the (5$_{\rm -4}$--4$_{\rm -4}$ E)
transition ($E_{\rm u}$=122 K). 
({\it d}): Same as (b) and (c) for the HDO(1$_{\rm 10}$--1$_{\rm 11}$) ($E_{\rm u}$=47 K) 
emission \citep{code05}. Since in this case the HPBW is 31$\arcsec$, the spectra 
taken at the (0$\arcsec$,+20$\arcsec$) and (0$\arcsec$,--20$\arcsec$) are shown.}
\end{figure*}

\begin{figure*}
\includegraphics[angle=-90,scale=0.9]{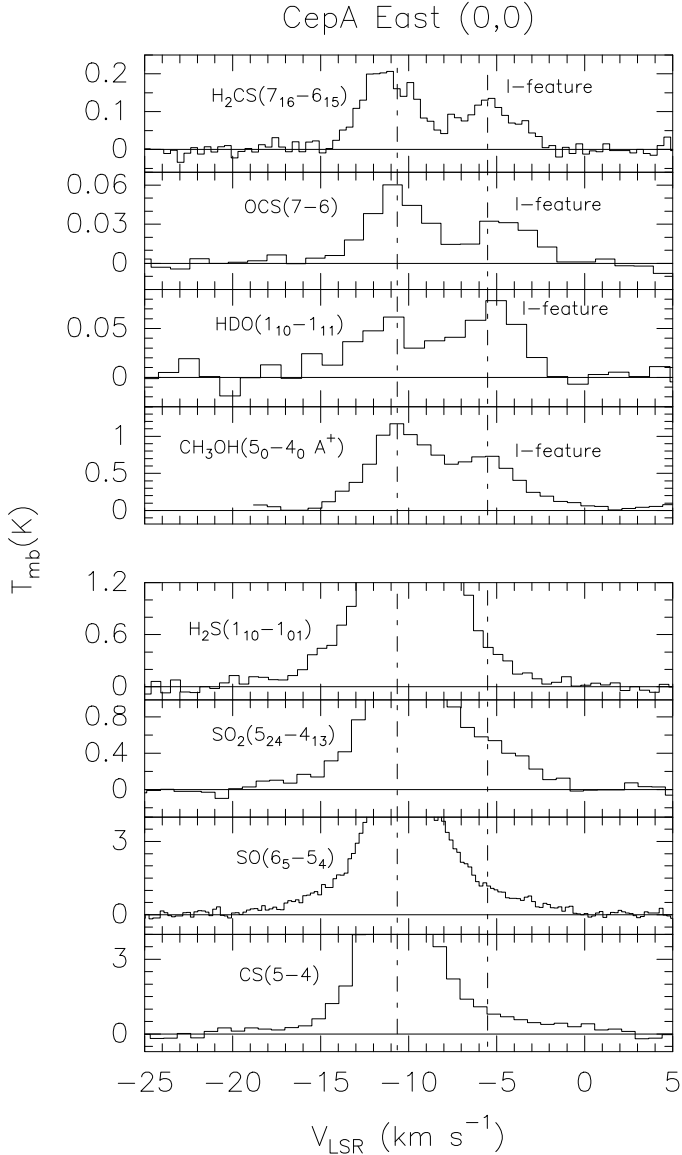}
\caption{Molecular line profiles observed towards CepA East \citep{code05}:
species and transitions are reported. Two different kinds of line profiles
are observed: (i) H$_2$S, SO$_2$, SO, and CS (lower panels) are associated
with extended wings spanning the whole range of outflow velocities, whereas (ii)
H$_2$CS, OCS, HDO, and CH$_3$OH (upper panels) are associated with wings and in addition
show a redshifted secondary peak at --5.5 km s$^{-1}$ (called I-feature in the text), 
well separated from the
ambient velocity (--10.7 km s$^{-1}$). The dashed lines stand for these two
velocities. The spectra in the lower panels have been truncated in order to focus
the attention on the spectral wings and to stress the absence of the
secondary peak clearly observed thanks to the H$_2$CS, OCS, HDO, and CH$_3$OH
emissions.}
\end{figure*}

\begin{figure*}
\includegraphics[angle=-90,scale=0.6]{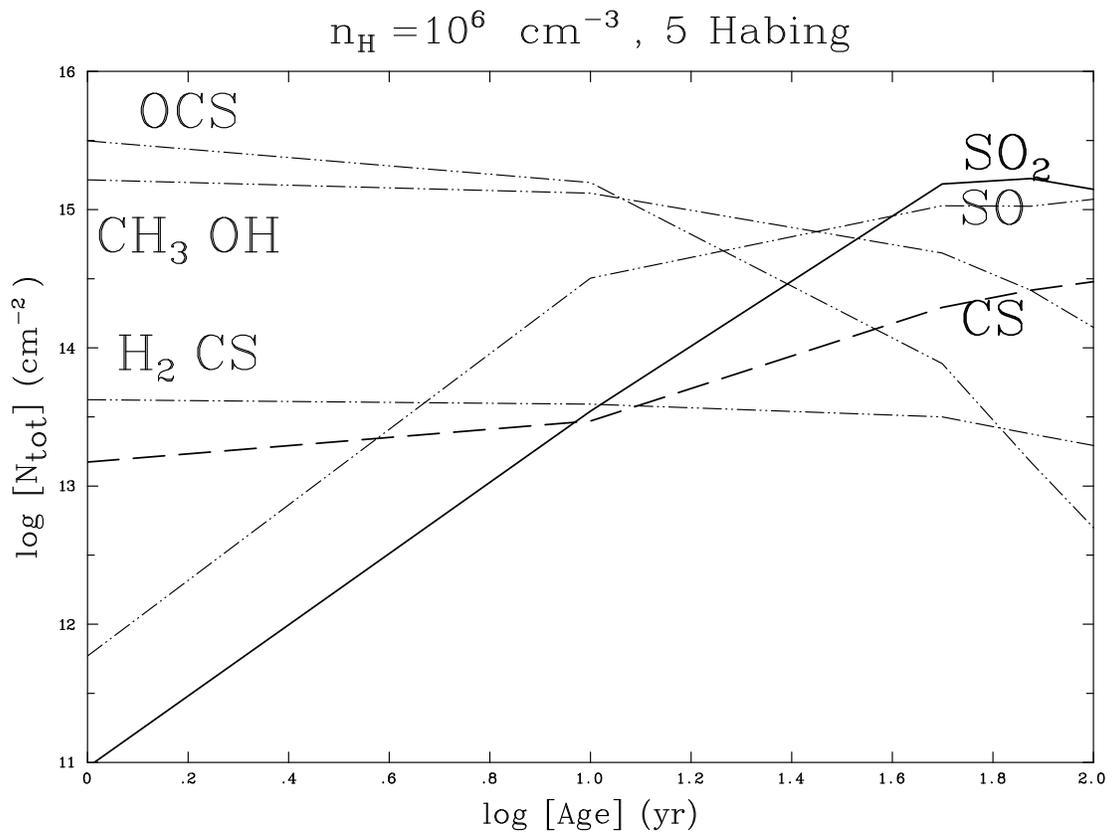}
\caption{Column densities of the double-peaked (H$_2$CS, OCS, CH$_3$OH)
and the single-peaked species (H$_2$S, SO$_2$, SO, CS) 
as a function of time for one of the best fit models of
molecular interfaces (see Sect. 3): the gas density is 10$^{6}$ cm$^{-3}$, while
the radiation field is 5 Habing. Note that the column density of H$_2$S is less 
than 10$^{11}$ cm$^{-2}$ and hence
it is undetectable.
Before evaporation, 100\% of the sulfur on the grains is in
the form of OCS.}
\end{figure*}

\end{document}